\documentclass[10pt]{iopart}
\usepackage{iopams}
\usepackage{setstack}
\usepackage[utf8]{inputenc}
\usepackage{cite}
\usepackage{siunitx}
\DeclareSIUnit\gauss{G}
\usepackage[colorlinks=true,linkcolor=blue,citecolor=blue,urlcolor=blue,pdfborder={0 0 0}]{hyperref}

\usepackage[color=blue!10!white]{todonotes}
\usepackage{braket}
\usetikzlibrary{calc,arrows,decorations.pathmorphing,intersections}
\usetikzlibrary{matrix}
\begin{document}
\title[Tailoring Rydberg interactions via Förster resonances]{Tailoring Rydberg interactions via Förster resonances: state combinations, hopping and angular dependence}
\author{Asaf Paris-Mandoki\footnote{a.paris-mandoki@physik.uni-stuttgart.de}, Hannes Gorniaczyk, Christoph Tresp, Ivan Mirgorodskiy, Sebastian Hofferberth\footnote{s.hofferberth@physik.uni-stuttgart.de}}
\address{5. Physikalisches Institut, University of Stuttgart, Germany}

\begin{abstract}
Förster resonances provide a highly flexible tool to tune both the strength and the angular shape of interactions between two Rydberg atoms. We give a detailed explanation about how Förster resonances can be found by searching through a large range of possible quantum number combinations. We apply our search method to $SS$, $SD$ and $DD$ pair states of $^{87}$Rb with principal quantum numbers from 30 to 100, taking into account the fine structure splitting of the Rydberg states. We find various strong resonances between atoms with a large difference in principal quantum numbers. We quantify the strength of these resonances by introducing a figure of merit $\tilde C_3$ which is independent of the magnetic quantum numbers and geometry to classify the resonances by interaction strength. We further predict to what extent excitation exchange is possible on different resonances and point out limitations of the coherent hopping process. Finally, we discuss the angular dependence of the dipole-dipole interaction and its tunability near resonances.

\end{abstract}


\vspace{2pc}

\maketitle

\newcommand{\Rb}{${}^{87}$Rb}
\newcommand{\vc}[1]{\vec{#1}\,} 
\newcommand{\uvc}{\hat} 

\newcommand{\bket}[2]{\left<#1|#2\right>}
\newcommand{\kbra}[2]{\left|#1\rangle\langle#2\right|}
\newcommand{\bopket}[3]{\left< #1 \left| #2 \right| #3 \right>}

\section{Introduction}
Ensembles of ultracold Rydberg atoms have proven to be a powerful tool for producing novel interacting quantum systems with wide-ranging applications from quantum information processing to quantum simulation~\cite{Saffman2010}. The underlying reason turning Rydberg gases into such a rich system is the strong dipole-dipole interaction between pairs of atoms~\cite{Cote2005,Shaffer2006,Walker2008,Pillet2010}. The strength and angular dependence of this interaction can be tailored by means of Förster resonances. These resonances occur whenever the energy of two coupled pair states is brought into degeneracy either by using electric~\cite{Safinya1981,Pillet2006,Raithel2008,Raithel2008b,Entin2010,Pfau2012,Pfau2012b,Weidemueller2013b,Marcassa2016} or microwave~\cite{Kachru1982,Martin2004,Martin2007} fields. With state-of-the-art laser and RF systems, a wide range of Rydberg states can be addressed. Experiments are exploiting the Rydberg blockade effect~\cite{Lukin2001} up to very high principal quantum numbers $n \approx 300$~\cite{Dunning2015,Pfau2013}. Single- or few-photon laser excitation enables addressing low angular-momentum states (typically $\ell=0,1,2$ or $3$), while additional RF fields or state mixing in static electric fields give access also to high $\ell$ states~\cite{Haroche2001,Dunning2003,Dunning2004,Dunning2008,Dunning2008b,Raithel2013,Hogan2015,Jia2015b}.

Of large interest from a technological point of view is to map the interaction between Rydberg states onto photons that coherently drive these transitions by means of Rydberg electromagnetically induced transparency (EIT). The resulting huge optical nonlinearities~\cite{Adams2010,Kuzmich2012,Vuletic2012,Vuletic2013b} have been employed to realize optical switches~\cite{Duerr2014}, transistors~\cite{Hofferberth2014,Rempe2014b} and enable imaging of single Rydberg atoms~\cite{Weidemueller2012,Lesanovsky2011}. The performance of these devices can be optimized by choice of quantum state combinations~\cite{Rempe2014b} and by tuning the levels into resonance by means of electric fields~\cite{Hofferberth2016}.
In particular, in such two-color experiments where different Rydberg states are addressed simultaneously, a large parameter space from which a choice can be made to enhance sought-after properties is opened up. Even more tunability is given in experiments with a mix of different atomic species~\cite{Saffman2015e}.
Furthermore, a strong dipole coupling results in a strong state mixture~\cite{Gallagher1998} and may be followed by state exchange betwen the atoms~\cite{Browaeys2014c,Cote2006,Buchleitner2011,Weidemueller2013,Browaeys2015}. This process is also found in photosynthesis as the underlying mechanism for energy transport~\cite{AspuruGuzik2008,Graham2007,Gregory2010}. However, when both spin-orbit coupling as well as Stark- and Zeeman splitting of all involved levels are taken into account, different resonances can exhibit very different features with a rich angular dependence~\cite{Browaeys2015b}. In this context, we provide generally applicable methods that quantify the distribution, strength, excitation exchange and angular dependence of Förster resonances. We exemplify these for the specific case of \Rb{}, a species used in many experiments.


\section{Interaction between Rydberg atoms near a Förster resonance}
\label{sec:interaction}

\begin{figure*}
\centering
\includegraphics{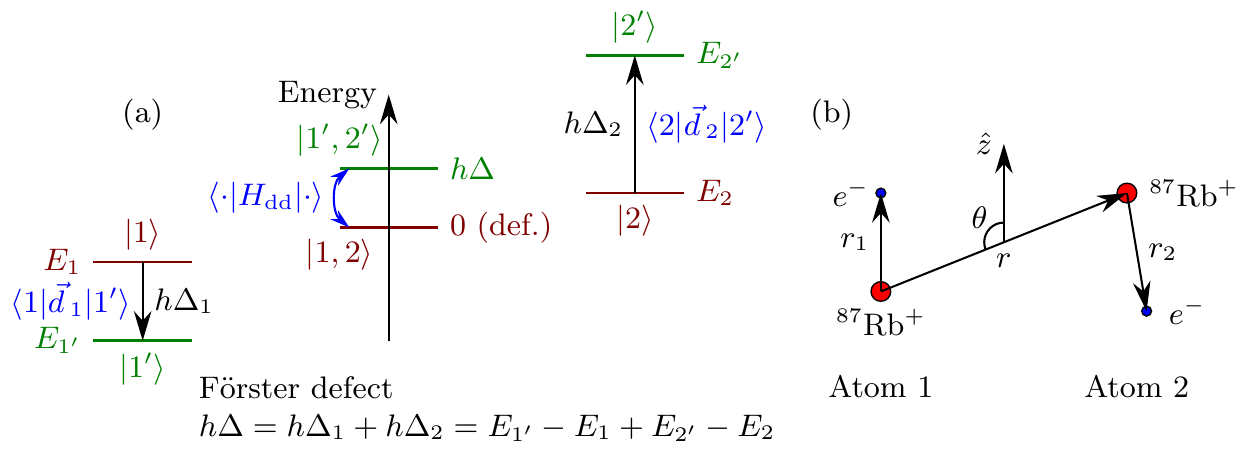}
\caption{ (a) The level structure of two Rydberg atoms, prepared in $\Ket{1}$ (left) and $\Ket{2}$ (right), affects the interaction strength between them by dipole-coupling to other states $\Ket{1'}$ and $\Ket{2'}$. For low energy differences (Förster defect) $\Delta$ of the pair states, the interaction is maximized. (b) Both the distance $r$ between the Rydberg atoms and the angle $\theta$ between this inter-atomic axis and the quantization axis $\hat z$ determine the interaction strength.}
\label{fig:interactionscheme}
\end{figure*}
We consider a pair of Rydberg atoms spatially separated by a distance $r$ as shown in Fig.~\ref{fig:interactionscheme}b. In the framework of the Born-Oppenheimer approximation, the position of the nuclei can be considered fixed with respect to the electronic motion and thus $r$ is treated as a scalar parameter and not an operator. For sufficiently large distance ($r > \langle r_1^2 \rangle^{1/2} + \langle r_2^2 \rangle^{1/2}$) that the individual Rydberg electron wavefunctions do not overlap, we solely need to consider the electrostatic interaction between the two separated charge distributions, which can be conveniently expressed as an infinite sum of interacting multipole terms~\cite{Dalgarno1966}. While various experiments have probed higher order interaction terms~\cite{Gould2008,Merkt2014,Loew2015}, we restrict our discussion to the dominating lowest order dipole-dipole term.
The interaction between two Rydberg atoms is then described by the total Hamiltonian
\[
H = H_1 \otimes \mathbb{I}_2 + \mathbb{I}_1 \otimes H_2 + H_\mathrm{dd},
\]
where $H_i$ is the single-atom Hamiltonian for atom $i$, operator $\mathbb{I}_i$ is the identity acting on the Hilbert space of atom $i$, and $H_\mathrm{dd}$ is the dipole-dipole interaction term.

As the problem involves two atoms, a sensible choice to describe the situation is a pair state basis, constructed as a tensor product of the individual atom bases. When $r\rightarrow\infty$, the energy of a pair state is simply the sum of the energies of the individual atoms. However, for smaller $r$ the interaction between the atoms will change this.

As an example, an atom that is \textit{initially} in a state $S$ can have dipole-dipole interaction with another atom in a $S'$ state even though eigenstates of the angular momentum magnitude operator $L^2$ have no dipole moment. The reason for this is that as the two atoms approach each other, dipole-coupling to close-by $\Ket{PP'}$ and $\Ket{P'P}$ pair states causes a state mixture of $S$ and $P$ which has a nonzero dipole moment. Thus, the proximity between the atoms induces a dipole moment in each and subsequently a dipole-dipole interaction. In fact, this mechanism is not restricted to only $S$-states but to any orbital angular momentum eigenstate.

Explicitly, the dipole-dipole interaction Hamiltonian is given by
\[
H_\mathrm{dd} = \frac{\vc{d}_1 \cdot \vc{d}_2-3(\uvc{r} \cdot \vc{d}_1)(\uvc{r} \cdot \vc{d}_2)}{r^3}.
\]
Here, $\vc{d}_1=e\vc{r}_1$ and $\vc{d}_2=e\vc{r}_2$ are the dipole moments of the individual atoms and $\uvc{r}$ is a unit vector in the direction of $\vc{r}$. Because of this interaction term of the Hamiltonian, a given state $\ket{1,2}$, will be coupled to other states $\ket{1',2'}$. This coupling between different states will produce off-diagonal terms in the full Hamiltonian. The interaction potential between atoms is then found by calculating the eigenvalues of the full Hamiltonian as function of the distance $r$.

To evaluate the matrix elements of the dipole-dipole operator, the single-atom dipole operators are first written in the spherical basis following the convention $d_{s\,\pm} = \mp (d_{s\,x}\pm i d_{s\,y})/\sqrt{2}$ for the operator acting on atom $s\in\lbrace 1,2\rbrace$. Thus, the Hamiltonian operator can be expanded as~\cite{Raithel2007},

\begin{eqnarray}\label{eq:angularHamiltonian}
H_\mathrm{dd} =& +\frac{d_{1z}d_{2z}(1-3\cos^2 \theta) - d_{1+}d_{2-} - d_{1-}d_{2+}}{r^3} \nonumber \\
&- \frac{3 \sin^2 \theta (d_{1+}d_{2+} + d_{1-} d_{2-} - d_{1+} d_{2-} - d_{1-} d_{2+})}{2 r^3}  \nonumber \\
&- \frac{3 \sin \theta \cos \theta (d_{1-} d_{2z} - d_{1+} d_{2z}+d_{1z}d_{2-} - d_{1z}d_{2+})}{\sqrt{2} r^3}.
\end{eqnarray}

In the following we consider Rydberg states including fine-structure splitting so that specific pair states are characterized by a total of eight quantum numbers $\ket{1,2}=\ket{n_1\ell_1j_1m_1,n_2\ell_2j_2m_2}$ and $\ket{1',2'}=\ket{n_1'\ell_1'j_1'm_1',n_2'\ell_2'j_2'm_2'}$. When evaluating the matrix elements  $\Braket{1,2|H_\mathrm{dd}|1',2'}$ it becomes clear that it is composed of a sum of terms of the form
\begin{equation}
\Braket{1,2|d_{1q_1}d_{2q_2}|1',2'} =\Braket{1 | d_{1q_1}| 1'} \Braket{2 | d_{2q_2}| 2'}.
\end{equation}
Each single-atom dipole matrix element $\Braket{k | d_{sq}| p}$  can be expressed as a product of two factors: a radial part $\tilde \mu_{k,p}$ that depends on all quantum numbers except the magnetic quantum numbers $m_k$, $m_p$, and an angular factor $\mathcal{C}_{k,p}^q$ which only depends on $j_k,m_k,j_p,m_p$, and the component index $q$ of the dipole operator. The single-atom matrix element is then written as
\begin{equation}\label{eq:reduction}
\Braket{k | d_{q}| p} =  \tilde\mu_{k,p}(n_k,\ell_k,j_k,n_p,\ell_p,j_p)  \mathcal{C}_{k,p}^q (j_k,\,m_k,\,j_p,\,m_p),
\end{equation}
where $\mathcal{C}_{k,p}^q$ is a coefficient given by
\begin{equation}\label{eq:ClebschGordans}
 \mathcal{C}_{k,p}^q= (-1)^{j_p - 1 + m_k}\left(\begin{array}{ccc} j_p& 1& j_k\\ m_p& q& -m_k\end{array}\right)
\end{equation}
with $(:::)$ denoting the Wigner 3-$j$ symbol. The radial factor is given by
\begin{eqnarray}
\label{eq:reduced_dipole_moment}
\tilde\mu_{k,p} &=& (-1)^{j_p+s+1}\sqrt{(2j_k+1)(2j_p+1)(2\ell_k+1)(2\ell_p+1)} \nonumber \\ 
&\times& \left\{
\begin{array}{ccc}
j_k & 1 & j_p \\
\ell_p & s & \ell_k
\end{array}\right\}
\left(
\begin{array}{ccc}
\ell_k & 1 & \ell_p \\
0 & 0 & 0
\end{array}\right)
\int_0^\infty R_k(r) er R_p(r) r^2\,dr 
\end{eqnarray}
where $\{:::\}$ is a Wigner 6-$j$ symbol and $R_s(r)$ is the radial wave function of state $s$.

Since the radial factor does not depend on $q$,  it is the same for all terms in a matrix element of the operator defined in eq.~\ref{eq:angularHamiltonian}  and therefore can be factored out:

\begin{eqnarray} \label{eq:angularReducedHamiltonian}
\hspace{-2.5cm}
 \Braket{1,2|H_\mathrm{dd}|1',2'} &=& \frac{\tilde\mu_{1,1'}\tilde\mu_{2,2'}}{r^3}  \biggl[ \left(\mathcal{C}_{1,1'}^{\pi}\mathcal{C}_{2,2'}^{\pi}(1-3\cos^2\theta ) - \mathcal{C}_{1,1'}^{\sigma^+}\mathcal{C}_{2,2'}^{\sigma^-} - \mathcal{C}_{1,1'}^{\sigma^-}\mathcal{C}_{2,2'}^{\sigma^+} \right) \\
 &&-\frac{3}{2} \sin ^2 \theta \left( \mathcal{C}_{1,1'}^{\sigma^+}\mathcal{C}_{2,2'}^{\sigma^+} +\mathcal{C}_{1,1'}^{\sigma^-}\mathcal{C}_{2,2'}^{\sigma^-}   -\mathcal{C}_{1,1'}^{\sigma^+}\mathcal{C}_{2,2'}^{\sigma^-}   -\mathcal{C}_{1,1'}^{\sigma^-}\mathcal{C}_{2,2'}^{\sigma^+}\right)  \nonumber \\
 && -\frac{3}{\sqrt{2}} \sin \theta \cos \theta \left( \mathcal{C}_{1,1'}^{\sigma^-}\mathcal{C}_{2,2'}^{\pi} - \mathcal{C}_{1,1'}^{\sigma^+}\mathcal{C}_{2,2'}^{\pi} + \mathcal{C}_{1,1'}^{\pi}\mathcal{C}_{2,2'}^{\sigma^-}   -\mathcal{C}_{1,1'}^{\pi}\mathcal{C}_{2,2'}^{\sigma^+}\right)\biggr] \nonumber \\
 &:=& \frac{\tilde\mu_{1,1'}\tilde\mu_{2,2'}}{r^3} A_{j_1\,m_1,j_2\,m_2,j_1'\,m_1',j_2'\, m_2'}(\theta) := \frac{C_3(\theta)}{r^3}. \label{eq:C3def}
\end{eqnarray}

The angular dependence is contained in the factor $A_{j_1\,m_1,j_2\,m_2,j_1'\,m_1',j_2'\, m_2'}(\theta)$ which depends only on the angular momentum quantum numbers and, for example, its structure will be common to describe the interaction between all pair states with the same numbers. In particular, this part of the Hamiltonian contains all information about the anisotropic character of the interaction. A more detailed discussion of this factor is carried out in section~\ref{sec:angular}.
On the other hand, the factor $\tilde\mu_{1,1'}\tilde\mu_{2,2'}$ may vary several orders of magnitude depending on the states used and it is this factor that determines the overall interaction strength given that the angular one is of order unity. An overview of the different possibilities of pair states, focusing on practical implications, is presented in sections~\ref{sec:overview} and \ref{sec:angular}. 




While the dipole-dipole Hamiltonian $H_\mathrm{dd}$ provides the coupling between different pair states, the atomic part of the Hamiltonian $H_1+H_2$ determines the initial energy difference $h\Delta$ between the specific pair states  as shown in Fig.~\ref{fig:interactionscheme}a. Although the separate energy levels for each of the atoms can be quite different, these levels can result  in a small Förster defect in the pair state basis.
If such a near-resonance exists the involved levels make up the dominant contribution to the dipole-dipole interaction. Therefore, the Hamiltonian which, in principle, is an infinite-dimensional matrix can be approximated by much smaller matrix where only near-resonant states are involved.

For a large Förster defect ($|h\Delta| \gg |C_3(\theta)/r^3|$), the dipole-dipole coupling can be understood as a second-order perturbation to the atomic Hamiltonian and the resulting interaction has an $\sim 1/r^6$ dependence~\cite{Ostrovsky2005,Cote2005,Walker2008}.
A so-called Förster resonance occurs when the energy of two coupled pair states is degenerate ($\Delta=0$ in Fig.~\ref{fig:interactionscheme}a). This degeneracy results in a resonant coupling between the states leading to a  $\sim 1/r^3$ dependence of the interaction.

In general, the near-to-resonance Hamiltonian includes all of the magnetic sub-levels. However, it is worth considering an example of a special case where the Hamiltonian matrix further simplifies. This example provides insight into the basic situations underlying other, more general circumstances. We consider an $\ket{S_{1/2},\uparrow,S'_{1/2},\uparrow}$ state separated by a small energy difference $h \Delta$ to the $\ket{P_{1/2},m_P,P'_{1/2},m_P'}$  and $\ket{P'_{1/2},m_P',P_{1/2},m_P}$ states.

For simplicity we consider the two atoms aligned along the quantization axis ($\theta=0$ in \ref{fig:interactionscheme}b). In this case, there are only two different nonzero off-diagonal entries in the Hamiltonian:

\begin{eqnarray*}
C_3/r^3 &= \Braket{S_{1/2},\uparrow,S'_{1/2},\uparrow|H_\mathrm{dd}|P_{1/2},\uparrow,P'_{1/2},\uparrow} =\tilde\mu_{1,1'}\tilde\mu_{2,2'}A(0)/r^3\\
C_3'/r^3 &= \Braket{S_{1/2},\uparrow,S'_{1/2},\uparrow|H_\mathrm{dd}|P'_{1/2},\uparrow,P_{1/2},\uparrow}  = \tilde\mu_{1,2'}\tilde\mu_{2,1'}A(0)/r^3.  \\
\end{eqnarray*}
The Hamiltonian then has the from
\begin{eqnarray}
H
&=\left(\begin{array}{cccc}
0	&	C_3/r^3	&	C_3'/r^3	&	0\\
C_3/r^3	&	h\Delta	&	0	&	C_3'/r^3\\
C_3'/r^3	&	0	&	h\Delta	&	C_3/r^3\\
0	&	C_3'/r^3	&	C_3/r^3	&	0
\end{array}\right)  \label{eq:C3matrix}                                                                                                                                                                                                                                                                                                                                                                                                                                                                  \end{eqnarray}
in the basis
\begin{equation}
 \mathcal{B} = \{ \Ket{S,S'},\Ket{P,P'},\Ket{P',P},\Ket{S',S}\} ,
\end{equation}
where the angular momentum numbers $J=1/2$ and $m_J=\,\uparrow$ have been omitted for brevity.
As a result of the dipolar coupling of the pair states, an atom in the initial state $\Ket{S_{1/2},S'_{1/2}}$ evolves into new eigenstates that can in general be written as
\begin{equation}
\Psi = \alpha_1 \Ket{S,S'} + \alpha_2 \Ket{P,P'} + \alpha_3 \Ket{P',P} + \alpha_4 \Ket{S',S}
\end{equation}
with $|\alpha_1|^2+|\alpha_2|^2+|\alpha_3|^2+|\alpha_4|^2 = 1$ for normalization. 

Note that the reversed states are not redundant but are essential for describing the excitation exchange between the two individual atoms. In particular, $|\alpha_4|^2$  gives the probability to measure the atoms in their flipped or $hopped$ state after some time. This aspect will be treated in more detail in subsection~\ref{subsec:hopping}.

One should keep in mind that such a restriction to a few dominant states fails at small distances. 
Due to the $r^{-3}$ dependence of the dipole coupling other pair states contribute significantly even if they are far detuned. The potentials can become so deep that the strong admixture can lead to a significant redistribution of Rydberg states~\cite{Gallagher1998,Pillet1998,Raithel2004,Tate2004,Shaffer2006} and the attraction can cause Penning ionization~\cite{Raithel2008a}.

\section{Electrically tuning the interaction strength}

\begin{figure*}
\centering
\includegraphics{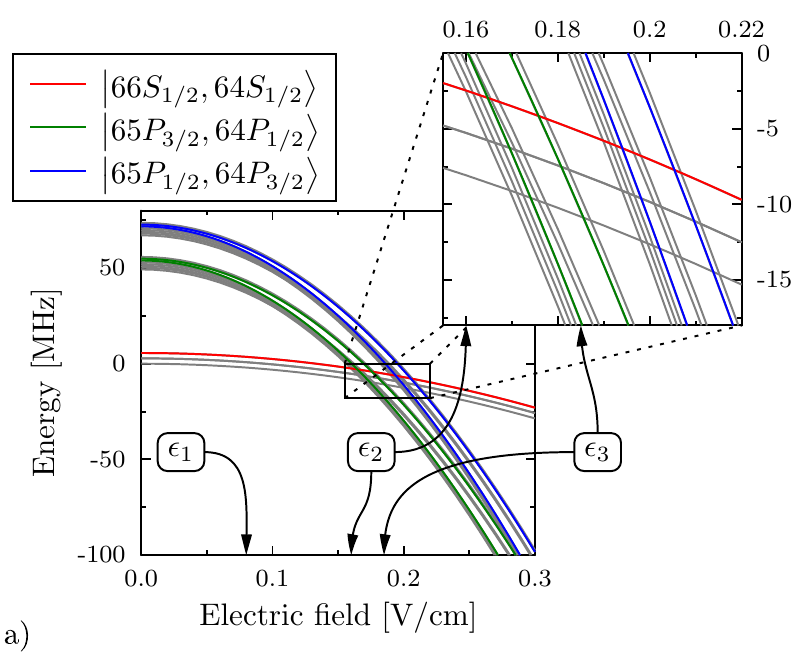}\includegraphics{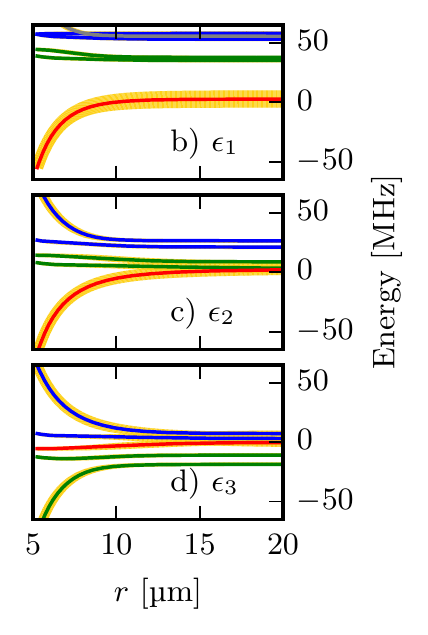}
\caption{The panels in a) show the Stark shift of Zeeman pair states for a magnetic field of $\SI{1}{\gauss}$ parallel to the electric field. The prepared state $m_1=m_2=1/2$ (red) couples at $\theta = 0$ to the green and blue lines. In the panels b), c) and d) the interaction of two Rydberg atoms is calculated by diagonalizing the dipole-dipole interaction Hamiltonian matrix at different electric fields. The basis used consist of $\num{14e3}$ coupled pair states. For large $r$, the lines correspond to the states in a) of the same color. We encode the overlap $|\braket{66S_{1/2},64S_{1/2}|\Psi_i}|^2$ between each of the lines $\Psi_i$ and the initial state in the opacity of the yellow stripe surrounding the lines to visualize the state mixing.}
\label{fig:potentials}
\end{figure*}

One way to bring pair states into degeneracy is to use an electric field. The different polarizabilities of Rydberg states allows tuning certain pair states into degeneracy at specific values of the field  as illustrated in Fig.~\ref{fig:potentials}a. Because of the multiple Zeeman sub-states, several Förster resonances are in close proximity when applying a nonzero magnetic field.

While off resonance the effective potential simply has an $\sim r^{-6}$ behavior as shown in Fig.~\ref{fig:potentials}b, on Förster resonance the situation is more complicated as is shown in Fig.~\ref{fig:potentials}c. Here the relevant potentials show an $\sim r^{-3}$ dependence.
As a result of the resonant dipole coupling, for any distance, the new eigenstates are superposition states with significant admixture from multiple unperturbed pair states.
Consequently, it is no longer possible to associate any potential curve with one asymptotic pair state $\ket{1,2}$. 
In particular a pair of atoms prepared in an unperturbed pair state will oscillate between the different eigenstates in the presence of interactions~\cite{Hofferberth2015}.

For the multi-resonance case (Fig.~\ref{fig:potentials}) contributions from different states can cancel each other and flatten the pair potential. This can be used to drastically reduce the Rydberg interaction~\cite{Hofferberth2016}. One has to keep in mind that the initial state will not be stationary and in particular  for small $r$ has a reduced overlap with the state corresponding to the flattened potential as shown by the yellow stripe in Fig.~\ref{fig:potentials}d.

\section{Overview of pair states}
\label{sec:overview}

\begin{figure*}
\centering
\includegraphics{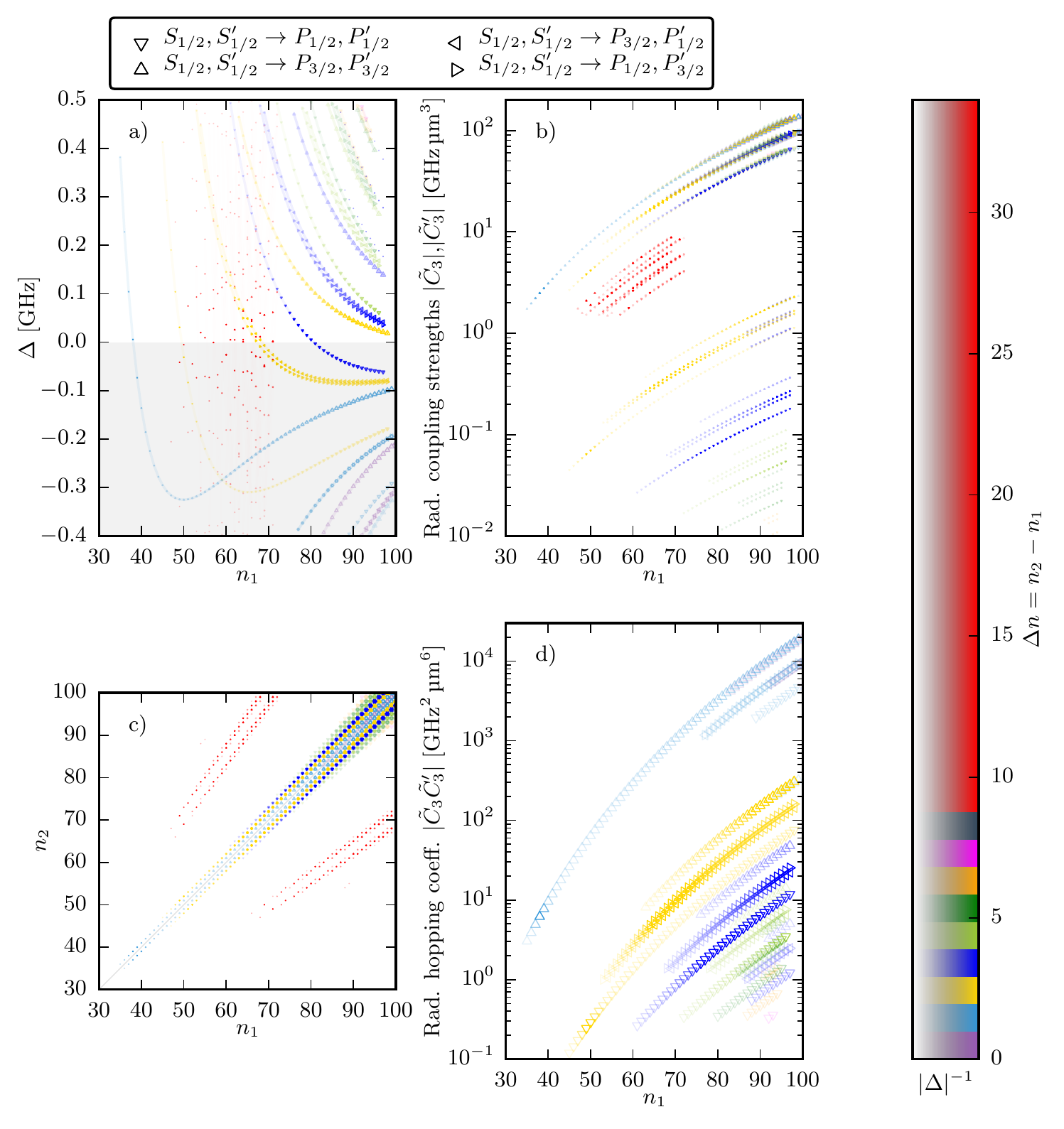}

\caption{Interaction properties between two Rydberg atoms initially in an $SS'$-state. Each state combination results in two data points in these figures, one corresponding to $\tilde C_3$ and one for $\tilde C_3'$. Plots are shown as a function of principal quantum number of the first atom $n_1$ while $n_2$ is encoded in the color of each data point. The opacity of the points represents the Förster defect, where a higher opacity is used for a smaller defect. In plots a), b), and c), the area of each point is proportional to $\tilde C_3$ or $\tilde C_3'$ accordingly. a) Förster defect as a function of $n_1$. Lines are a guide to the eye and relate similar states which only differ in $n_1$. The $\Delta<0$ region is grayed out because it is not accessible using a static electric field. b) Radial coupling constants as a function of $n_1$. c) Overview of all $SS'$-states. The visibility of the points depends on both having a small defect and a significant radial coupling constant. d) Hopping coefficient as a function of $n_1$. }
\label{fig:SS}
\end{figure*}

\begin{figure*}
\centering
\includegraphics{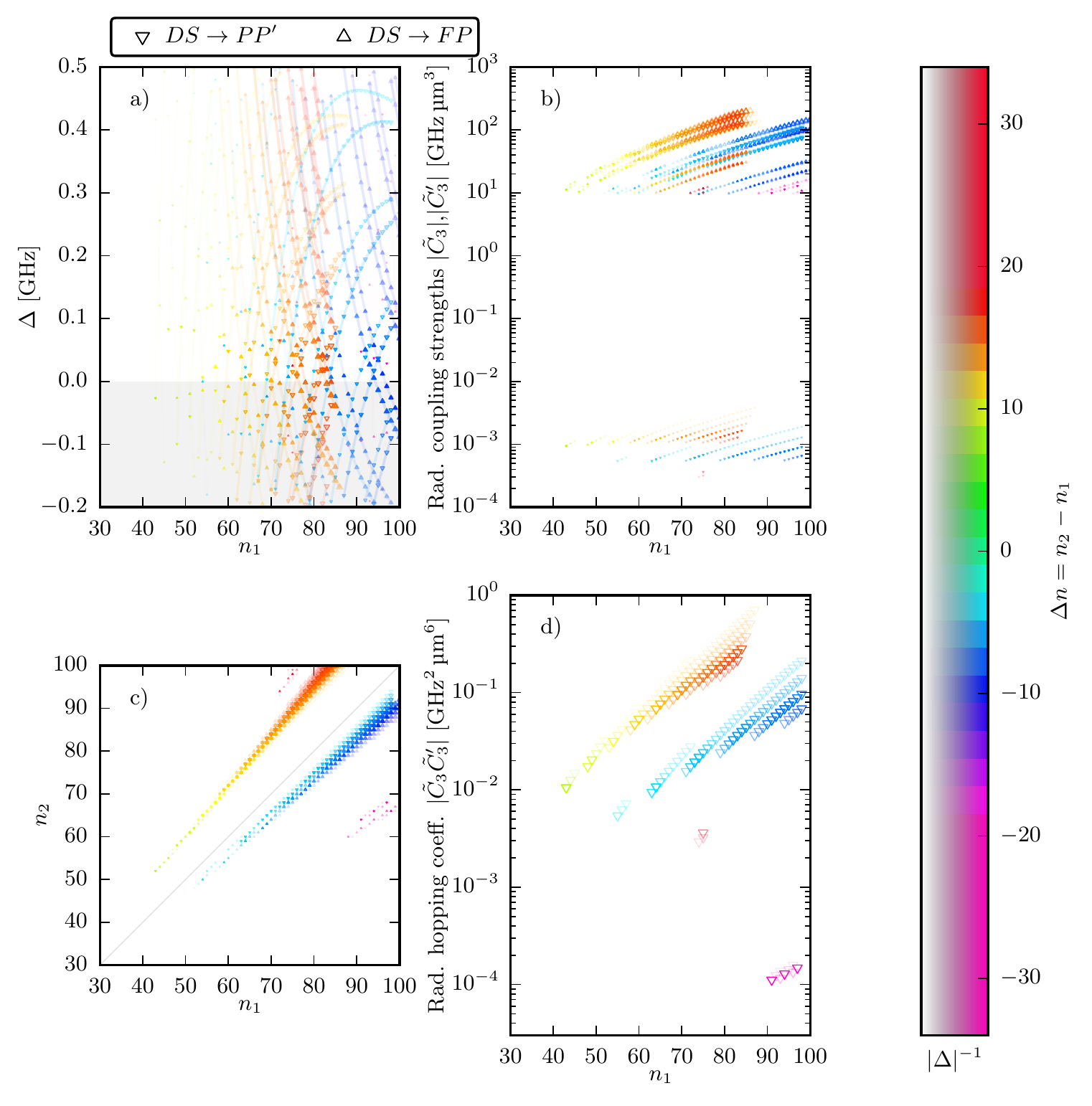}
\caption{Interaction properties between two Rydberg atoms initially in an $SD$-state. Each state combination results in two data points in these figures, one corresponding to $\tilde C_3$ and one for $\tilde C_3'$. Plots are shown as a function of principal quantum number of the first atom $n_1$ while $n_2$ is encoded in the color of each data point. The opacity of the points represents the Förster defect, where a higher opacity is used for a smaller defect. In plots a), b), and c), the area of each point is proportional to $\tilde C_3$ or $\tilde C_3'$ accordingly. a) Förster defect as a function of $n_1$. Lines are a guide to the eye and relate similar states which only differ in $n_1$. The $\Delta<0$ region is grayed out because it is not accessible using a static electric field. b) Radial coupling constants as a function of $n_1$. c) Overview of all $SD$-states. The visibility of the points depends on both having a small defect and a significant radial coupling constant. d) Hopping coefficient as a function of $n_1$.}
\label{fig:SD}
\end{figure*}

\begin{figure*}
\centering
\includegraphics{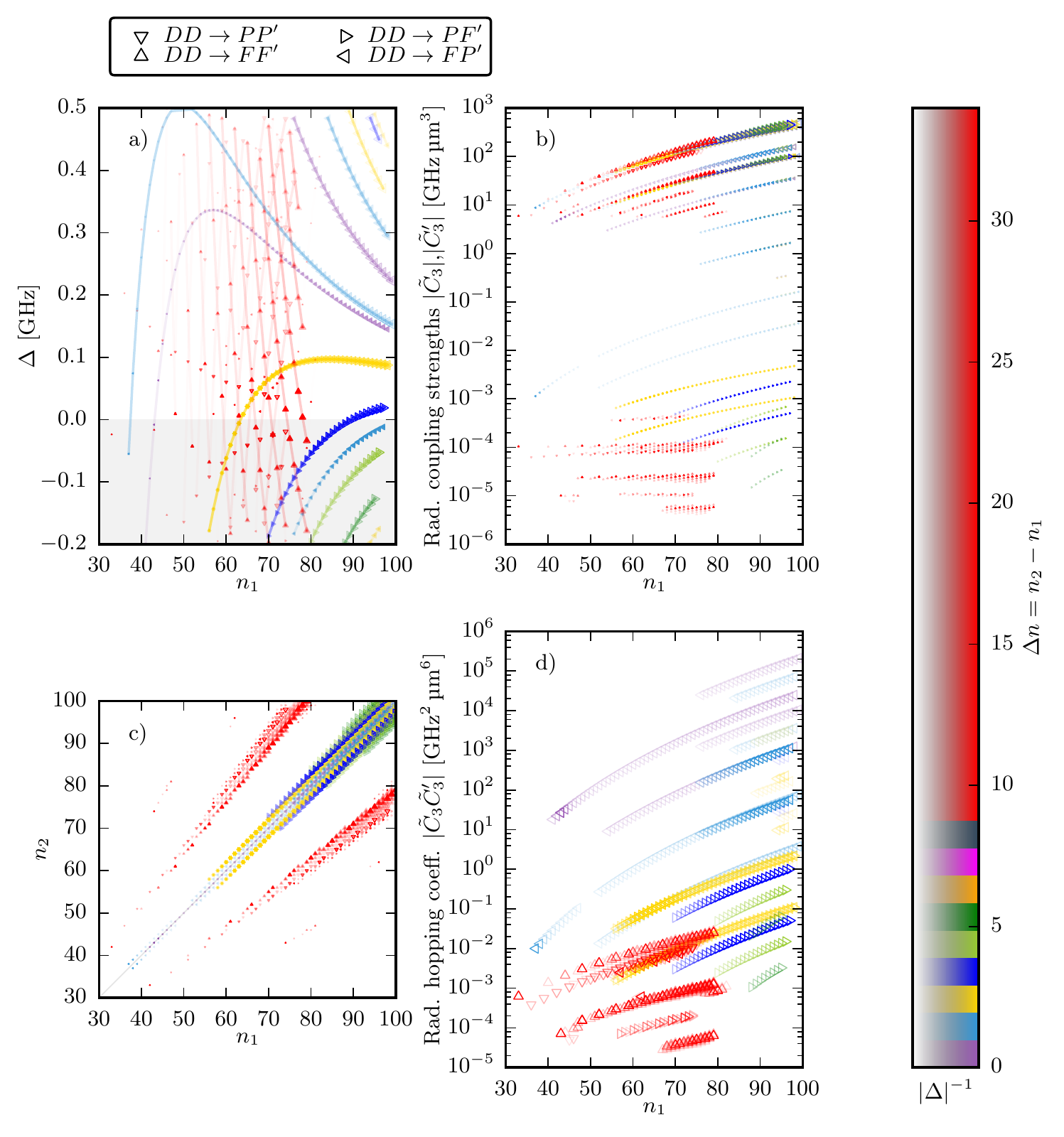}
\caption{Interaction properties between two Rydberg atoms initially in an $DD'$-state. Each state combination results in two data points in these figures, one corresponding to $\tilde C_3$ and one for $\tilde C_3'$. Plots are shown as a function of principal quantum number of the first atom $n_1$ while $n_2$ is encoded in the color of each data point. The opacity of the points represents the Förster defect, where a higher opacity is used for a smaller defect. In plots a), b), and c), the area of each point is proportional to $\tilde C_3$ or $\tilde C_3'$ accordingly. a) Förster defect as a function of $n_1$. Lines are a guide to the eye and relate similar states which only differ in $n_1$. The $\Delta<0$ region is grayed out because it is not accessible using a static electric field. b) Radial coupling constants as a function of $n_1$. c) Overview of all $DD'$-states. The visibility of the points depends on both having a small defect and a significant radial coupling constant. d) Hopping coefficient as a function of $n_1$.}
\label{fig:DD}
\end{figure*}

The aim of this section is to classify a range of experimentally accessible pair states according to their interaction properties. For this intent, we focus on the radial factors between an initial state $\ket{1,2}$ coupled through $H_\mathrm{dd}$ to the primed states $\ket{1',2'}$ and $\ket{2',1'}$ defining the \emph{radial coupling factors}
\begin{eqnarray*}
\tilde C_3 &= \tilde\mu_{1,1'}\tilde\mu_{2,2'}\\
\tilde C_3' &= \tilde\mu_{1,2'}\tilde\mu_{2,1'},
\end{eqnarray*}
where the energy difference (Förster defect) between $\ket{1,2}$ and $\ket{1',2'},\ket{2',1'}$ is $h \Delta$.
Keeping in mind that to precisely determine the interaction potential the angular factors should be accounted for as well, we focus only on the quantities $\tilde C_3$ and $\tilde C_3'$ since they give an overall measure of the dependence of the interaction strength on the choice of initial states.





The vastness of states to be considered, which includes all possible initial states and all possible primed states, is first reduced by limiting the range of the 12 quantum numbers.
Here we restrict our initial state combinations to two \Rb{} atoms in states with principal quantum numbers between 30 and 100. Furthermore, only initial pair states composed of different $S$-states, different $D$-states and $SD$-state combinations are considered. 
For a given initial state, the selection rules for the dipole operator drastically reduce the set of primed quantum numbers, for instance, in the sets $\Ket{n_1S_{1/2},n_2S_{1/2}}$ we include $\ell_1',\ell_2' = P$ with $J_1',J_2' = 1/2,3/2$ only. For $\Ket{n_1S_{1/2},n_2D_{5/2}}$ we include $\ell_1' = P$ with $J_1' = 1/2,3/2,$ and $\ell_2' = P, J_2' = 3/2$ or $\ell_2' = F, J_2'=5/2,7/2$ only.

Finally, we reduce the primed states to those with zero-field Förster defect smaller than $\Delta E_{cut}=\SI{500}{\mega\hertz}$ since these contribute the most to the interaction.

The energy of individual Rydberg levels is calculated using the Rydberg constant from reference~\cite{Fortagh2011} and quantum defects from~\cite{Fortagh2011,Gallagher2003,Gallagher2006}.
For obtaining the radial dipole moments we use  radial wave functions calculated numerically using a model potential for the Rydberg valence electron~\cite{Dalgarno1994}.

\subsection{Interaction strength}
\label{subsec:interactionstrength}

In figures \ref{fig:SS},\ref{fig:SD} and \ref{fig:DD} results are presented for initial pair states of type $\Ket{\ell_1=S,\ell_2=S}$, $\Ket{\ell_1=S,\ell_2=D}$ and $\Ket{\ell_1=D,\ell_2=D}$ respectively. For each set, four plots are shown. In this subsection we focus on plots marked with a), b) and c).  The Förster defect plots a) allow us to identify the most easily accessible Förster resonances.

As can be observed in the figures, a wide range of possibilities are available. Several close-to resonance pair states can be found with varying degree of strength and for different $\Delta n=n_2-n_1$. As shown most clearly in c) plots,  the strongest resonances for $SS'$ and $SD$ state combinations are for small differences in principal quantum numbers $\Delta n$ as found in previous works~\cite{Walker2008,Saffman2015e}. However, there are surprisingly strong resonances in $n_1D_{5/2},n_2D_{5/2}$ state combinations with $n_2-n_1 \gg 1$. For instance, we found a strong zero field resonance
\[
\Ket{78D_{5/2}, 99D_{5/2}} \longleftrightarrow \Ket{77F_{7/2}97F_{7/2}},
\]
where $\tilde C_3 = \SI{206}{\giga\hertz\,\micro\metre ^3}$, $\tilde C_3' \approx \SI{0}{\giga\hertz\,\micro\metre ^3}$ and $\Delta =  \SI{3}{\mega\hertz}$, which is approximately as strong as the strongest resonance with small $\Delta n$
\[
\Ket{78D_{5/2}, 80D_{5/2}} \longleftrightarrow \Ket{79P_{3/2}79F_{7/2}},
\]
for which  $\tilde C_3 = \SI{184}{\giga\hertz\,\micro\metre ^3}$, $\tilde C_3' \approx \SI{0}{\giga\hertz\,\micro\metre ^3}$ and $\Delta =  \SI{92}{\mega\hertz}$.
The latter has to be tuned to Förster resonance by an electric field due to the large $\Delta$.

These resonances with a large difference in principal quantum numbers greatly increase the available options for accessing a Förster resonance at zero electric field if no electric field control is possible or if the Stark admixture of other states which comes along with finite electric fields is not an option. Secondly, these combinations are particularly interesting for Rydberg transistors~\cite{Hofferberth2014,Rempe2014b,Hofferberth2016} and interaction-enhanced imaging~\cite{Weidemueller2012,Lesanovsky2011}, where the desired situation is a maximized interaction between two different states ($78D,99D$), and minimized interaction between atoms in the low state ($78D,78D$).

\subsection{Excitation exchange}
\label{subsec:hopping}


If $\tilde C_3, \tilde C_3'>0$, the flip-flop (\textit{hopping}) process might be possible on resonance ($h\Delta=0$) as:

\begin{tikzpicture}
  \matrix (m) [matrix of math nodes, row sep=0em, column sep=3em]{
    & \ket{1',2'} & \\
    \ket{1,2} & & \ket{2,1}  \\
    & \ket{2',1'} & \\};\\
   \path (m-2-1) -- node[sloped] (text) {$\tilde C_3$} (m-1-2);
   \draw[<-] (m-2-1) edge (text);
   \draw[->] (text) edge (m-1-2);
   \path (m-3-2) -- node[sloped] (text) {$\tilde C_3$} (m-2-3);
   \draw[<-] (m-3-2) edge (text);
   \draw[->] (text) edge (m-2-3);
   \path (m-1-2) -- node[sloped] (text) {$\tilde C_3'$} (m-2-3);
   \draw[<-] (m-1-2) edge (text);
   \draw[->] (text) edge (m-2-3);
   \path (m-2-1) -- node[sloped] (text) {$\tilde C_3'$} (m-3-2);
   \draw[<-] (m-2-1) edge (text);
   \draw[->] (text) edge (m-3-2);
\end{tikzpicture}

\noindent
where the result is that the two atoms exchange their internal state as a consequence of dipole-dipole interaction~\cite{Zoubi2015}. It is worth noting that, if the two initial single atom states are not fully spin-polarized in the stretched states of equal signs, this process may not swap the Zeeman states. For example, two $S$-states that undergo this process
$
 \Ket{n_1S_{1/2} \downarrow,n_2S_{1/2}\uparrow} \nonumber
 \stackrel{\tilde C_3}{\longleftrightarrow}\Ket{n_1'P_{1/2}\downarrow,n_2'P_{1/2}\uparrow}\label{eq:noHopping}
 \stackrel{\tilde C_3'}{\longleftrightarrow}\Ket{n_2S_{1/2}\downarrow,n_1S_{1/2}\uparrow}\label{eq:mchangingforsterprocess}\nonumber
$
, do not end up in the fully flipped state because the  $\pi$ coupling is the only non-zero term in eqn.~(\ref{eq:angularReducedHamiltonian}).
This simple example shows that while some  of the quantum numbers are exchanged, not the full quantum state is exchanged. In the case where both atoms are in $m=\,\uparrow$ or both are in $m=\,\downarrow$, an exchange of the full quantum state is only guaranteed at $\theta = 0$. For non-zero angles, $m_1+m_2$ is not conserved, as can be directly seen in eqn.~(\ref{eq:angularReducedHamiltonian}), and the coupling caused by interaction does not result in a closed transition.
If we consider the interaction in the full Zeeman Basis, i.e., we calculate $C_3,C_3'$ instead of $\tilde C_3,\tilde C_3'$, then, $C_3, C_3' \neq 0$ is a sufficient condition for the hopping. However, as described above, because of the different interaction channels for the magnetic quantum numbers, the quantum state may not flip completely. To sum up, the experimental geometry and level structure has to be carefully examined to judge if a \textit{true} state exchange is possible.

In order to quantify the tendency of a pair state to flip into another state we consider a Hamiltonian with a form as in eq.~\ref{eq:C3matrix} on resonance ($\Delta=0$). The flipping probability after a short time $t$ is given by
\begin{eqnarray*}
\braket{2,1|e^{-i H t/\hbar}|1,2} &\approx& \braket{2,1|1-i H t/\hbar- H^2 t^2/2 \hbar^2|1,2}  \\
 &=& -t^2/2 \hbar^2 \braket{2,1| H^2 |1,2} \\
 &=& C_3C_3'(-t^2 /\hbar^2 r^6).
\end{eqnarray*}
This motivates the definition of a \emph{hopping coefficient} given by $C_3 C_3'$. Again, this term includes the angular and radial factors and we can define a \emph{radial hopping coefficient} by $\tilde C_3 \tilde C_3'$. We plot this coefficient for $SS$, $SD$ and $DD$ combinations in figures~\ref{fig:SS}d, \ref{fig:SD}d and \ref{fig:DD}d respectively. It is striking that this quantity may vary over ten orders of magnitude depending on the choice of states.

When choosing a Rydberg pair state for performing a specific experiment, the coupling strength plots b) and the radial hopping coefficient plots d) should be read together. For example, for an energy transfer experiment, it is desirable to have strong coupling together with strong hopping. However, for an experiment involving interaction-enhanced imaging where hopping is not desirable, a strongly coupled pair state with reduced hopping can be chosen.

\section{Angular dependence of interaction}
\label{sec:angular}

\begin{figure*}
\centering
\includegraphics{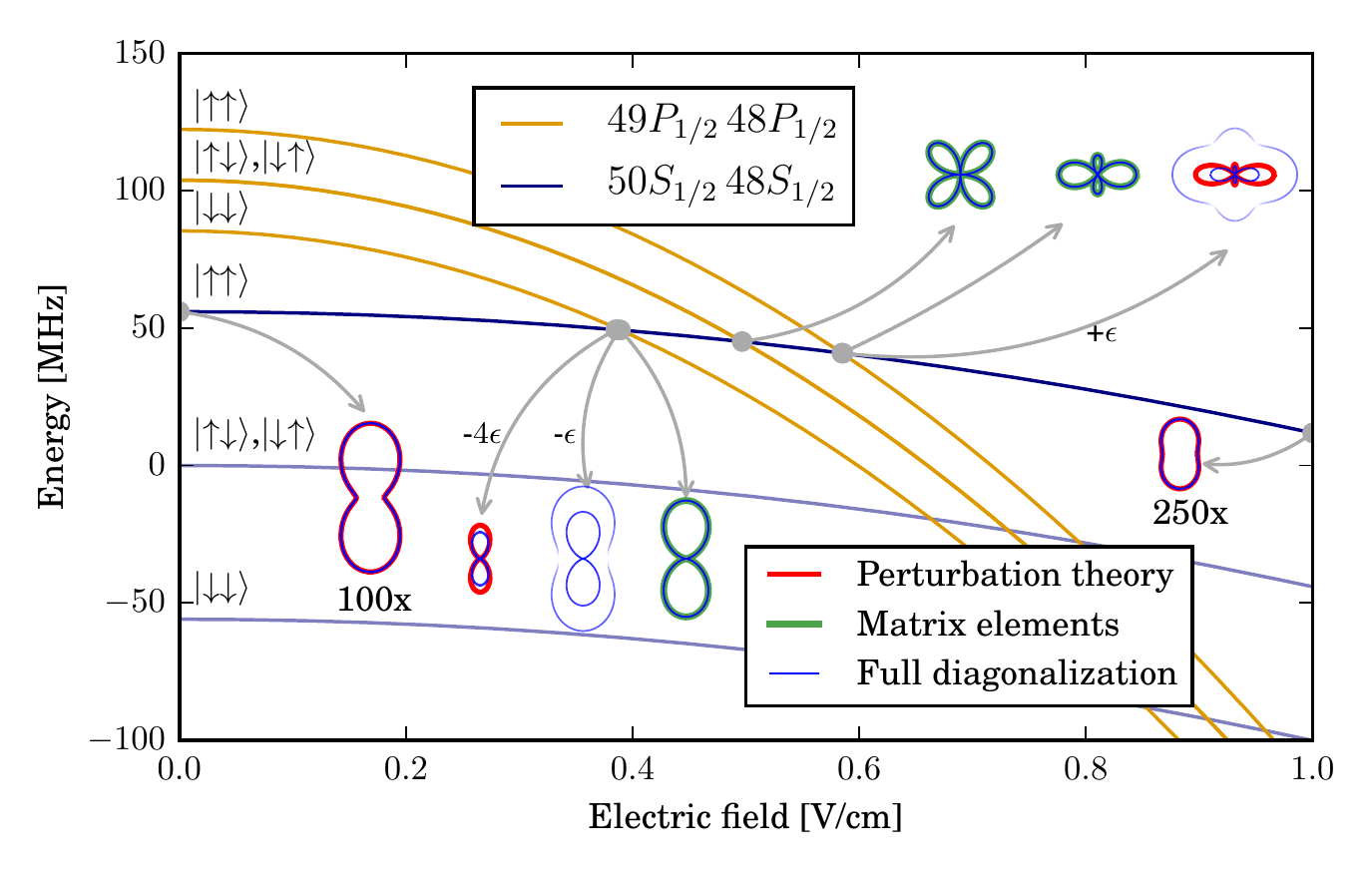}
\caption{Angular dependence of the dipole-dipole interaction. The pair state energies of the $\ket{50 S_{1/2} \uparrow, 48S_{1/2} \uparrow}$ and  $\ket{49 P_{1/2} m_1, 48P_{1/2} m_2}$ states are shown as a function of electric field subject to a magnetic field of \SI{20}{\gauss} parallel to the electric field. Strong $\theta$-dependent dipole coupling between the pair states leads to a set of new eigenstates with $\theta$-dependent eigenenergies. The green polar plots (horizontal axis is $\theta = 0$) show the respective coupling matrix elements. The eigenenergies are shown in the blue polar plots and the projection of new eigenstates onto the initial state is encoded in the line opacity. The exact diagonalization coincides with the matrix elements on the resonances and with the perturbative calculation far from resonance. As the Förster defect increases (leftmost and rightmost polar plots), the interaction energy is reduced, so we enlarge the plots by a factor of 50 and 250 as indicated. The plots next to the matrix element plots are slightly off resonance by an amount indicated in the arrow, where $\epsilon = \SI{1}{\milli\volt/\centi\meter}$. The dipole coupling was calculated at a distance of $r=\SI{1.5}{\micro\meter}$.}
\label{fig:angular}
\end{figure*}

So far, only the radial factors of the matrix element in eq.~\ref{eq:C3matrix} have been discussed. However in general, the angular factors must be accounted for to get a complete picture of the interaction~\cite{Shaffer2011}. In this section the angular factors  are discussed paying special attention to the resulting angular dependence of the interaction in the vicinity of a Förster resonance where highly anisotropic interaction can occur.

In order to understand the angular behavior of the interaction, we consider the case where $\tilde{C}_3' \approx 0$ while $\tilde{C}_3 \neq 0$ in which the hopping dynamics described in subsection~\ref{subsec:hopping} are not present. Furthermore, the magnetic sub-levels do not couple to each other under this condition and the Hamiltonian matrix has the general form
\[
H = \left(
\begin{array}{cc}
0 & \vc{C}_3^T \\
\vc{C}_3 & \mathrm{diag}(h\vc{\Delta})

\end{array}
\right)
\]
where the first element of the basis is the initial state $\ket{1,2}$ with fixed values of $j_1$,$m_1$,$j_2$ and $m_2$ and the following ones are all the magnetic substates of energetically close pair state $\ket{1',2',p}$. Here, $p$ indexes all the possible values of $m_1'$ and $m_2'$. Also, the components of $\vc{C}_3$ are $C_3^p = \bopket{1,2}{H_\mathrm{dd}}{1',2',p}$ which are calculated using eq.~\ref{eq:C3def}. Furthermore, $\mathrm{diag}(h\vc{\Delta})$ is a diagonal matrix with entries given by $h\Delta_p$ which is the energy of the level $\ket{1',2',p}$ with the energy origin set to that of the $\ket{1,2}$ state. It is worth noting that the $\theta$-dependence is completely contained in $\vc{C}_3$.

For the sake of concreteness we consider an initial state  $\ket{S^1_{1/2},\uparrow,S^2_{1/2},\uparrow}$ and a neighboring  $\ket{P^{1'}_{1/2},m_P^{1'},P^{2'}_{1/2},m_P^{2'}}$  pair state as shown in Fig.~\ref{fig:angular}. For this case, the Hamiltonian matrix takes the form
\[
H = \left(
\begin{array}{ccccc}
0 & C_3^{\downarrow\downarrow}/r^3 & C_3^{\downarrow\uparrow}/r^3 & C_3^{\uparrow\downarrow}/r^3 & C_3^{\uparrow\uparrow}/r^3 \\
C_3^{\downarrow\downarrow}/r^3 & h\Delta_{\downarrow\downarrow} & 0 & 0 & 0 \\
C_3^{\downarrow\uparrow}/r^3  & 0 & h\Delta_{\downarrow\uparrow} & 0 & 0 \\
C_3^{\uparrow\downarrow}/r^3  & 0 & 0 & h\Delta_{\uparrow\downarrow} & 0 \\
C_3^{\uparrow\uparrow}/r^3  & 0 & 0 & 0 & h\Delta_{\uparrow\uparrow}
\end{array}
\right),
\]

where the $C_3$ coefficients depend on $\theta$ and the energy defects depend on the magnitude of the electric and magnetic fields. Here we consider only the case where the state mixing caused by the electric field is negligible.

Applying an external magnetic field causes the Förster defects $h\Delta_p$ of each of the magnetic sub-states to be different. As a result, several closely-spaced Förster resonances  appear. The resulting eigenvalues of the Hamiltonian, which give the interaction potential, will have contributions of varying importance that arise from the various angular factors and therefore, the shape of the interaction will depend on the external electric field.

On resonance with a specific state $p_0$, the angular dependence will be predominantly determined by the $C_3^{p_0}(\theta)$ factor. By changing the electric field, the angular dependence of the interaction will also change because the weight of the different angular contributions is also modified.

However, as is the case with the radial dependence of the interaction, exactly on resonance there is no single eigenvalue that can be identified as the interaction potential for atoms in the initial pair state. In fact, there are two such eigenvalues $\pm C_3^{p_0}(\theta)/r^3$ whose corresponding eigenvectors 
\[
\frac{1}{\sqrt{2}}\left(\ket{1,2}\pm\ket{1',2',p_0}\right),
\]
have a significant overlap with the initial pair state as illustrated in Fig.~\ref{fig:angular}. Here, both eigenvalues are shown~(blue) but overlap each other, as well as with the matrix elements~(green). On the other hand, in the vicinity of the resonance, these two eigenvalues become different as seen in the plots indicated with $-4\epsilon$, $-\epsilon$, and $+\epsilon$. While exactly on resonance the two eigenstates corresponding to these eigenvalues have a $50\%$ overlap with the initial pair state, away from resonance the overlap of one of the eigenstates increases while the other decreases. The result of this is that away from resonance one eigenvector can be well identified with the initial pair state as the case shown in the polar plots corresponding to \SI{0}{\volt/\centi\meter} and \SI{1}{\volt/\centi\meter}.

Far from resonance (when $|C_3^p(\theta)/r^3|\ll |h \Delta_p| $), second order perturbation theory can be used and the interaction potential is given by
\begin{equation}
V(r,\theta)=-\sum_{p} \frac{(C_3^p(\theta))^2}{r^6 h\Delta_p},
\label{eq:second_order_angular}
\end{equation}
where it is evident how the different defects $\Delta_p$ give different weighting to the various angular factors resulting in an E-field dependent anisotropy of the interaction. The perturbation theory results are shown in Fig.~\ref{fig:angular}. While far from resonance they have a very good agreement with the eigenvalues, close to resonance this approximation breaks down. The shape of the interaction changes the most in the vicinity of the Förster resonance where the perturbative treatment is not valid. 

It is clear from Fig.~\ref{fig:angular} that by adjusting the electric field, the anisotropy of the interaction can be tuned from a side-by-side interaction (resonance at lowest E-field) to a head-to-tail interaction (resonance at highest E-field). These multi-state resonances thus greatly increase the tunability of the angular shape of the Rydberg interaction additionally to the significant boost of the interaction strength.

\section{Conclusion}
\label{sec:conclusion}

By comparing a wide range of possible pair state combinations, a set of promising pair states, suited for different kinds of experiments were found. These pair states are classified according to $\tilde C_3$ which is independent of the magnetic quantum number and geometry. Using this quantity as a figure of merit, strongly interacting pair states with large differences in principal quantum number were found.

The energy-transfer dynamics were also considered and a hopping coefficient to quantify hopping rates was proposed. Using this measure, states that exhibit hopping dynamics ranging over several orders of magnitude were obtained.
Furthermore, the angular dependence of the interaction and its tunability via electric and magnetic fields was discussed. It was shown that can be tuned, for example, from being a head-to-tail to side-by-side interaction.

\section{Acknowledgments}

We thank Sebastian Weber for calculation of Rydberg potentials and Przemyslaw Bienias for fruitful discussions. This work is funded by the German Research Foundation through Emmy-Noether-grant HO 4787/1-1, within the SFB/TRR21 and by RiSC grant 33-7533.-30-10/37/1 from the Ministry of Science, Research and the Arts of Baden-Württemberg. H.G.\ acknowledges support from the Carl-Zeiss Foundation. 

\section{References}

\bibliographystyle{iopart-num}
\bibliography{biblio2}

\end{document}